\documentclass[12pt]{article}
\usepackage{amssymb}
\usepackage{amsfonts,amssymb,amsmath,amsthm}
\usepackage{epsfig}
\usepackage{slashed}
\usepackage{graphicx}
\usepackage{mathrsfs}
\usepackage{bbm}
\def\and{{\rm and}}

\def\t{\tau}
\def\om{\omega}

\def\th{\theta}

\begin{document}
\renewcommand{\thefootnote}{\fnsymbol{footnote}}
\begin{titlepage}
\begin{flushright}
USTC-ICTS-10-22
\end{flushright}

\vspace{10mm}
\begin{center}
{\Large\bf Tunneling of massive particles from noncommutative inspired Schwarzschild black hole}
\vspace{16mm}

{{\large Yan-Gang Miao${}^{1,2,}$\footnote{\em E-mail:
miaoyg@nankai.edu.cn}, Zhao Xue${}^{1,}$\footnote{\em E-mail:
illidanpimbury@mail.nankai.edu.cn}
and Shao-Jun Zhang${}^{1,}$}\footnote{\em E-mail: sjzhang@mail.nankai.edu.cn}\\

\vspace{6mm}
${}^{1}${\em School of Physics, Nankai University, Tianjin 300071, \\
People's Republic of China}

\vspace{3mm}
${}^{2}${\em Interdisciplinary Center for Theoretical Study,\\
University of Science and Technology of China, Hefei, Anhui 230026,\\
People's Republic of China}}

\end{center}

\vspace{10mm}
\centerline{{\bf{Abstract}}}
\vspace{6mm}
We apply the generalization of the Parikh-Wilczek method to the tunneling of massive particles from noncommutative inspired Schwarzschild black holes.
By deriving the equation of radial motion of the tunneling particle directly, we calculate the emission rate which is shown to be dependent
on the noncommutative parameter besides the energy and mass of the tunneling particle. After equating the emission rate to the Boltzmann factor,
we obtain the modified Hawking temperature which relates to the noncommutativity and recovers the standard Hawking temperature in the
commutative limit. We also discuss the entropy of the noncommutative inspired Schwarzschild black hole and its difference
after and before a massive particle's emission.
\vskip 20pt
PACS Number(s): 04.70.-s, 04.70.Dy, 11.10.Nx

\end{titlepage}
\newpage

\pagenumbering{arabic}

\section{Introduction}
Black holes are an incredible outcome of the Einstein's theory of gravity. They were thought that no matter inside could escape
and thus they were invisible from outside.
In 1970's, Hawking startled the physical community by proving~\cite{Hawking} that the black holes are not really
black. They can radiate like a black body with the ``Hawking temperature", $T_H = \frac{\kappa}{2\pi}$,
where $\kappa$ is the surface gravity of black holes. Although Hawking first suggested the heuristic picture of the radiation as tunneling,
his calculation was completely based on quantum field theory in the curved spacetime, which is independent of a tunneling process.

An intuitional method was proposed at the end of 1990's by Parikh and Wilczek~\cite{Wilczek99} in which
the radiation is dealt with as tunneling of particles. It is suggested in the method that the barrier is created by the tunneling particle itself.
With the WKB approximation, the emission rate is calculated semiclassically. In terms of the equality of the emission rate and the Boltzmann factor,
the Hawking temperature is then derived.
The results show that the emission spectrum is not precisely thermal but its leading order coincides with the standard Hawking radiation,
i.e. the black body spectrum. However, it is only investigated in ref.~\cite{Wilczek99} that
massless and uncharged particles tunnel from spherically symmetric black holes, such as the Schwarzschild black hole
and Reissner-Nordstr\"{o}m black hole. When the tunneling of a massive and charged particle is considered,
the problem that the trajectory of the tunneling particle is not null geodesic has to be confronted and thus the equation of
motion of the massive charged particle has to be found out. In ref.~\cite{ZhengZhao2005},
the concepts of phase velocity and group velocity are introduced
in order to derive the equation of radial motion for a massive and charged tunneling particle, and the equation of radial motion is obtained
through equating the radial velocity of the tunneling particle to the phase velocity. In addition,
a quite different method based on the Hamilton-Jacobi equation
is provided in refs.~\cite{Srinivasan,Angheben:2005,Kerner,Banerjee}.
This method is not involved in the trajectory problem, and can thus be applied to tunneling particles with mass and/or charge naturally.
Recently, the Parikh-Wilczek method has been generalized~\cite{Mxz2010} to the tunneling particle with mass and charge
in terms of strictly deriving its equation of radial motion.
The corresponding results show that the emission rate depends on
the mass and charge of the tunneling particle, and that the radiation temperature relates to the charge but
equals the standard Hawking temperature in the limit of zero charge.

On the other hand, the idea of noncommutative (NC) spacetimes has been revived since the publication
of the Seiberg and Witten's work~\cite{Seiberg1999}. This idea is originally expected to remove the divergence in quantum field theory~\cite{Snyder}
and later expected to be an indispensable ingredient for quantization of gravity~\cite{DFR}.
Till now a large amount of papers have been devoted to the construction of quantum field theories on noncommutative spacetimes,
for a review, see ref.~\cite{NCQFT}. Gravity theories on noncommutative spacetimes have extensively been studied,
for instance, see the review article~\cite{Szabo2006}, and black hole solutions have also been obtained~\cite{NCbh}.
Most of the works related to NC spacetimes are based on the ``star-product approach" in which
the noncommutativity of spacetimes is encoded through an ordinary product in the
noncommutative $C^{\ast}$-algebra of Weyl operators to a noncommutative star-product of functions
in commutative $C^{\ast}$-algebra.
Within this framework one calculates only order by order in noncommutative parameters
and therefore loses the nonlocality of noncommutative theories. Nevertheless, the so-called ``coordinate coherent states approach"
has been proposed~\cite{Smailagic and Spallucci} from a different point of view for the study of noncommutative quantum mechanics
and quantum field theory.
The models established in quantum field theory with such an approach are consistent with the Lorentz invariance, unitarity and UV-finiteness.
The main idea of the approach is that the point matter distribution should be depicted by a Gaussian function rather
than a Dirac delta function. In terms of this smeared matter distribution,
the NC inspired Schwarzschild black hole solution is provided~\cite{Nicolini2005} under the assumption that the Einstein's equation is not influenced
by the noncommutativity. The remarkable property of the solution is that there exits a minimal mass $M_0$
under which no horizon can form and therefore there will be a remnant after the Hawking radiation, which may probably solve
the ``information loss paradox". Based on ref.~\cite{Nicolini2005}, a lot of efforts have been devoted to extend the solution to higher
dimensions and to charged and rotating black holes, such as for the NC inspired Schwarzschild black hole solution in higher dimensions~\cite{Rizzo2006},
for the NC inspired Reissner-Nordstr\"{o}m black hole solution~\cite{Nicolini2006,Nicolini2008},
and for the NC inspired Kerr(-Newmann) black hole solution~\cite{Smailagic2010,Nicolini2010}. Furthermore,
the thermodynamics of the various NC inspired black holes mentioned above has been analyzed in accordance with the Parikh-Wilczek method or
Hamilton-Jacobi method, in which only the tunneling of massless particles is considered~\cite{NCbhmassless}.
In addition, the massive particle's tunneling from NC inspired black holes is discussed in ref.~\cite{NCbhmassive}
in terms of  the method given by ref.~\cite{ZhengZhao2005}.

Our motivation is obvious from the above summary on both the aspect of the Hawking radiation as a tunneling process and the aspect of the
tunneling from NC inspired black holes, that is,
we follow our generalization of the Parikh-Wilczek method~\cite{Mxz2010} to
study the tunneling process of massive particles from the NC inspired Schwarzschild black hole.
In the next section, we give a brief introduction to
the NC inspired Schwarzschild black hole. In section 3, starting from the Lagrangian of the tunneling particle, we derive its equation of radial motion
directly and then calculate its emission rate. After equating the emission rate to the Boltzmann factor, we therefore obtain the modified Hawking
temperature which relates to the noncommutativity of spacetimes. In the following section we discuss the entropy of the
NC inspired Schwarzschild black hole and its difference after and before a massive particle's emission. Finally,
we make a conclusion in section 5.

\section{Noncommutative inspired Schwarzschild black hole}
Consider a noncommutative spacetime where $\theta$ denotes the noncommutative parameter.
In the coordinate coherent states approach~\cite{Smailagic and Spallucci},
a particle (black hole) with mass $M$ is smeared over a region with width $\sqrt{\theta}$ and is described by a Gaussian function
rather than a Dirac delta function as follows,
\begin{eqnarray}
\rho_{\th}(r)=\frac{M}{\left(4\pi\th\right)^{\frac{3}{2}}}e^{-\frac{r^2}{4\th}}.
\end{eqnarray}
If the Einstein's equation is not influenced, the Schwarzschild-like solution of the Einstein's equation
with the matter source mentioned above takes the form~\cite{Nicolini2005},
\begin{eqnarray}
ds^2=-\left(1-\frac{2M_\th}{r}\right)dt^2+\left(1-\frac{2M_\th}{r}\right)^{-1}dr^2+r^2d\Omega^2,\label{Schwsolution}
\end{eqnarray}
where the parameter $M_{\th}$ with the dimension of mass satisfies the following formula,
\begin{eqnarray}
M_{\th}(r)=\int^r_0\rho_\th(r^{\prime})4\pi{r^{\prime}}^2dr^{\prime}=\frac{2M}{\sqrt \pi} \gamma\left(\frac{3}{2},\frac{r^2}{4\th}\right).
\end{eqnarray}
The gamma function in the above equation is defined to be $\gamma (a,b)= \int_0^b \frac{d t}{t} \ t^a e^{-t}$.

Here we make a comparison between the coordinate coherent state approach and the star-product approach.
For the former which is argued~\cite{Vorosproduct,NCbhmassless} to be equivalent to the so-called Voros-product (a specific star-product),
the effect of noncommutativity only on the matter is considered while the gravity is left untack. Such a treatment results in a smeared ``point mass"
described by a Gaussian function (see eq.~(1)), but it maintains the usual Einstein equation unchanged.
For the latter, the noncommutative effect on both the matter and gravity is considered,
and therefore it is in general hard to find exact solutions to the noncommutative Einstein equation.
Only in some very special cases can the nonperturbative solutions be found,
such as that discussed in ref.~\cite{Twistedbh} (mentioned by ref.~\cite{AGE}).
Based on the twisted noncommutative gravity theory~\cite{TDG},
the nonperturbative solutions are derived under some specific compatibility conditions of metrics and twists.
There is no proof that the coordinate coherent state approach yields a solution of the noncommutative Einstein equations,
such as that proposed in ref.~\cite{TDG} which works under the star-product approach.
Comparatively, it is not so hard to find exact solutions and to study the thermodynamics of various noncommutative inspired black holes
under the framework of the coordinate coherent state approach.
In some sense, the coordinate coherent state approach might be regarded as an effective way at least from a phenomenological viewpoint.
As a whole, the both approaches corresponding to different star-products
have their own merits in light of the present studies and more researches on various
black holes are needed if one tries to distinguish a better one from the two.

The solution (eq.~(\ref{Schwsolution}))
describes a spacetime interpolating a de Sitter spacetime near the origin and a Schwarzschild spacetime far away from the origin. Taking the limit
$r\rightarrow \infty$ in the gamma function, one can see that the parameter $M_\th$ is just the total mass $M$ of the source. In this limit,
the solution (eq.~(\ref{Schwsolution})) coincides with the classical vacuum Schwarzschild solution. Alternatively, we can get this result by
taking the limit
$\th\rightarrow 0$, which presents the degeneration from the noncommutative to the commutative case. Furthermore, the square of line element
has no singularity at the origin, which can be shown by the Ricci scalar near the origin,
\begin{equation}
R (0) = \frac{4 M}{\sqrt{\pi} \th^{\frac{3}{2}}}.
\end{equation}

If the equation $g_{00} (r_H) = 0$ has solutions, there exist horizons. That is, the horizon $r_H$ is determined by the following relation,
\begin{equation}
r_H = 2 M_\th (r_H) = \frac{4M}{\sqrt{\pi}} \gamma\left(\frac{3}{2}, \frac{r_H^2}{4 \th}\right).\label{horizonequation}
\end{equation}
However, it cannot be solved analytically. The numerical analysis shows~\cite{Nicolini2005} that
only when the total mass $M$ is larger than the minimal value, $M_0 \sim 1.9 \sqrt{\th}$,
can the horizons be formed\footnote{The authors of ref.~\cite{Nozari2007} use modified dispersion relation to analyze the thermodynamics
of the noncommutative inspired Schwarzschild black hole and give some interesting results
which show that the remnant mass may be larger than $M_0$.}.
This means that there will be a remnant after the end of Hawking radiation.

Due to the fact that $\gamma$ function tends to unit quickly, the ordinary Schwarzschild horizon $r^{\prime}_H=2M$ is a good approximation.
To have a more precise approximation to reflect the noncommutative effect, one substitutes $r^{\prime}_H=2M$ into eq.~(\ref{horizonequation})
and thus has the relation,
\begin{eqnarray}
r_H=2M_\th\left(r^{\prime}_H\right)=2M\left({\rm erf}\left(\frac{M}{\sqrt{\th}}\right)-\frac{2M}{\sqrt{\pi\th}}e^{-\frac{M^2}{\th}}\right),
\label{horizon}
\end{eqnarray}
where the Gaussian error function is defined as ${\rm erf}(x)=\frac{2}{\sqrt \pi}\int^x_0e^{-t^2}dt$.
When we repeat this procedure, $r_H$ will get closer and closer to the true value.
In this paper we only consider the first iterative approximation given by eq.~(\ref{horizon}).

\section{Quantum tunneling of massive particles}
Now we turn to analyze the tunneling process of massive particles by following the idea of the Parikh-Wilczek method~\cite{Wilczek99}
together with its generalization~\cite{Mxz2010}.
The first thing we have to do is to derive the equation of radial motion of the tunneling particle.
As we know, when a free massive particle with mass $m$ moves along a time-like geodesics, its trajectory is determined by the Lagrangian,
\begin{eqnarray}
2 \mathcal{L}= m g_{\mu\nu}\frac{dx^\mu}{d\t}\frac{dx^\nu}{d\t},\label{lagrangian}
\end{eqnarray}
where the parameter $\t$ is the proper time. The choice of this parameter provides an extra equation,
\begin{eqnarray}
g_{\mu\nu}\dot{x^\mu}\dot{x^\nu}=-1,\label{time-like geodesic}
\end{eqnarray}
where a dot stands for the derivative with respect to $\tau$.
Substituting the metric (eq.~(\ref{Schwsolution})) into the Lagrangian (eq.~(\ref{lagrangian})), we have
\begin{eqnarray}
2\mathcal{L}=-m\left(1-\frac{2M_\th}{r}\right)\left({\frac{dt}{d\t}}\right)^2+\left(1-\frac{2M_\th}{r}\right)^{-1}\left({\frac{dr}{d\t}}\right)^2.
\label{lagrangian2}
\end{eqnarray}
From the Lagrangian we know that the canonical coordinate $t$ is a cyclic coordinate. As a result, the corresponding canonical momentum,
i.e. the particle's energy is conserved,
\begin{eqnarray}
-p_t=-\frac{\partial \mathcal{L}}{\partial \dot{t}}=m\left(1-\frac{2M_\th}{r}\right)\dot{t} \equiv \omega = {\rm const}.,\label{energy}
\end{eqnarray}
where the minus sign before $p_t$ presents due to the positivity of the energy $\omega$ of the tunneling particle.
Substituting eq.~(\ref{energy}) into eq.~(\ref{time-like geodesic}), we obtain
\begin{eqnarray}
\dot{r}= \pm m^{-1}\sqrt{\omega^2-m^2\left(1-\frac{2M_\th}{r}\right)}\, ,\label{radial}
\end{eqnarray}
where the upper (lower) sign in eq.~(\ref{radial}) corresponds to an outgoing (ingoing) geodesics.
Using eqs.~(\ref{energy}) and (\ref{radial}) we finally derive the equation of time-like geodesics outgoing along the radial direction,
\begin{eqnarray}
\frac{dr}{dt}=\frac{\dot{r}}{\dot{t}}=\frac{1}{\omega}\left(1-\frac{2M_\th}{r}\right)\sqrt{\omega^2-m^2\left(1-\frac{2M_\th}{r}\right)}.\label{dot r}
\end{eqnarray}

In order to investigate the Hawking radiation as a tunneling process, we have to transform the coordinates into the ones which have a good behavior at
the horizon. As done in ref.~\cite{Wilczek99}, we choose the Painlev\'e coordinates~\cite{Painleve1921} which can be acquired
by the following transformation,
\begin{eqnarray}
dt_p=dt+\frac{\sqrt{1-g_{00}}}{g_{00}}dr,\label{transformation}
\end{eqnarray}
where $t_p$ denotes the Painlev\'e time and the other components of coordinates keep unchanged.
Under this coordinate transformation, the square of line element (eq.~(\ref{Schwsolution})) becomes
\begin{eqnarray}
ds^2=-\left(1-\frac{2M_\th}{r}\right)dt^2_p+dr^2+2\sqrt{\frac{2M_\th}{r}}dt_pdr+r^2d\Omega^2.
\end{eqnarray}
Correspondingly, the equation of time-like geodesics (eq.~(\ref{dot r})) in the Painlev\'e coordinates takes the form,
\begin{eqnarray}
\frac{dr}{dt_p}=\left(1-\frac{2M_\th}{r}\right)\left(\frac{\omega}{\sqrt{\omega^2-m^2\left(1-\frac{2M_\th}{r}\right)}}
+\sqrt{\frac{2M_\th}{r}}\right)^{-1}.
\end{eqnarray}

In accordance with the WKB method, we have to get the imaginary part of the action of the tunneling particle for the calculation of the tunneling rate.
This imaginary part emerges from the integral of the radial coordinate if the corresponding
integrand has a singularity, and it is defined~\cite{Wilczek99} for the tunneling particle which crosses the horizon outwards
from $r_{\rm in}$ to $r_{\rm out}$ as follows,
\begin{eqnarray}
{\rm Im} S &\equiv& {\rm Im}\int^{r_{\rm out}}_{r_{\rm in}} p_r d r = {\rm Im} \int^{r_{\rm out}}_{r_{\rm in}} \int^{p_r}_0 d \tilde{p}_r d r
= {\rm Im} \int^{r_{\rm out}}_{r_{\rm in}} \int^{\omega}_{m} \frac{d H}{\frac{d r}{d t_p}} d r\nonumber\\
&=& - {\rm Im} \int^{r_{\rm out}}_{r_{\rm in}} \int^{\omega}_{m} \frac{d \tilde{\omega}}{\frac{d r}{d t_p}} d r,\label{Im action}
\end{eqnarray}
where the Hamilton equation $\frac{d r}{d t_p} = \frac{d H}{d \tilde{p}_r}$ has been utilized,
and the minus sign appears due to the relation~\cite{Wilczek99}:
$H=M-\tilde{\omega}$.
As explained in ref.~\cite{Mxz2010},
the lower bound of the energy is $m$ instead of
zero because the tunneling particle is massive, which leads to a modification to the emission rate.
When we consider the self-gravitation of the tunneling particle, we have to make the replacement
$M_\th(M)\rightarrow M_\th(M-\tilde{\omega})$ in eq.~(\ref{horizon}) and eq.~(\ref{Im action}). Moreover,
with eq.~(\ref{horizonequation}) we give the initial and final positions of the tunneling particle as follows,
\begin{eqnarray}
r_{\rm in} &=& \frac{4M}{\sqrt{\pi}} \gamma\left(\frac{3}{2}, \frac{r^2_{\rm in}}{4\th}\right),\\
r_{\rm out} &=& \frac{4\left(M-\omega\right)}{\sqrt{\pi}} \gamma\left(\frac{3}{2}, \frac{r^2_{\rm out}}{4\th}\right).
\end{eqnarray}

At present we exchange the order of integration in eq.~(\ref{Im action}) and pick out the $r$ integral,
\begin{eqnarray*}
& &{\rm Im} \int^{r_{\rm out}}_{r_{\rm in}}\left(\frac{dr}{dt_p}\right)^{-1}dr \nonumber \\
&=&{\rm Im} \int^{r_{\rm out}}_{r_{\rm in}}\left(1-\frac{2M_\th \left(M-\tilde{\omega}\right)}{r}\right)^{-1}
\left(\frac{\tilde{\omega}}{\sqrt{\tilde{\omega}^2-m^2\left(1-\frac{2M_\th \left(M-\tilde{\omega}\right)}{r}\right)}}
+\sqrt{\frac{2M_\th \left(M-\tilde{\omega}\right)}{r}}\right)dr.
\end{eqnarray*}
There is a pole at $r = 2M(M-\tilde{\omega})$ in the integrand. Integrating $r$ by deforming the contour and making use of the residue method
and eq.~(\ref{horizon}), we obtain
\begin{eqnarray}
{\rm Im} \int^{r_{\rm out}}_{r_{\rm in}}\left(\frac{dr}{dt_p}\right)^{-1}dr= - 4\pi M_\th \left(M-\tilde{\omega}\right).
\end{eqnarray}
Next we complete the energy integral in eq.~(\ref{Im action}) and at last work out the imaginary part of the action,
\begin{eqnarray}
{\rm Im} S&=&4\pi\int^{\omega}_mM_\th(M-\tilde{\omega})d\tilde{\omega}\nonumber\\
&=&-2\pi\left[(M-\omega)^2-\frac{3}{2}\th \right] {\rm erf}\left(\frac{M-\omega}{\sqrt\th}\right)
-6\sqrt{\pi\th}\,(M-\omega)\,e^{-\frac{(M-\omega)^2}{\th}}\nonumber\\
&&+2\pi\left[(M-m)^2-\frac{3}{2}\th\right]{\rm erf}\left(\frac{M-m}{\sqrt\th}\right)+6\sqrt{\pi\th}\,(M-m)\,e^{-\frac{(M-m)^2}{\th}}. \label{imaction}
\end{eqnarray}

The tunneling probability takes the form,
\begin{eqnarray}
\Gamma \sim e^{-2 {\rm Im} S}\sim e^{\Delta S_{\rm BH}},\label{entropy}
\end{eqnarray}
where $\Delta S_{\rm BH}$ is the difference of black hole entropy after and before the Hawking radiation.
From eqs.~(\ref{imaction}) and (\ref{entropy}) one can see that the emission rate depends not only on the energy of the tunneling particle
but also on its mass $m$ and the noncommutative parameter $\th$, and the dependence of $\th$ will lead to a modified Hawking temperature.
Expanding ${\rm Im} S$ with respect to $\omega$ and $m$ to their first orders, we get
\begin{equation}
{\rm Im} S=4\pi M \left[{\rm erf}\left(\frac{M}{\sqrt{\th}}\right) -\frac{2M}{\sqrt{\pi\th}} e^{-\frac{M^2}{\th}}\right](\omega-m)
+ \mathcal{O} (\omega,m).\label{first order}
\end{equation}
Equating the tunneling probability $\Gamma \sim e^{- 2 {\rm Im} S}$ to the Boltzmann factor $e^{- \frac{\omega}{T_H}}$,
we can deduce the modified Hawking temperature from eqs.~(\ref{entropy}) and (\ref{first order}),
\begin{equation}
T_H=\frac{1}{8\pi M \left[{\rm erf}\left(\frac{M}{\sqrt{\th}}\right) -\frac{2M}{\sqrt{\pi\th}} e^{-\frac{M^2}{\th}}\right]},\label{temperature}
\end{equation}
which is valid only when $M$ is larger than $M_0$, the mass of the extreme black hole.
When substituting eq.~(\ref{horizon}) into the above equation, we obtain a simple relation between the modified temperature
and the event horizon: $T_H=\frac{1}{4\pi r_H}$, which takes the same form as that of the ordinary Schwarzschild black hole.
However, we note that
$r_H$ is just the first iterative approximation for the NC inspired
Schwarzschild black hole while the same symbol stands for the exact value of event horizon
for the ordinary Schwarzschild black hole.
In the commutative limit $\th\rightarrow 0$, eq.~(\ref{temperature}) recovers the standard Hawking temperature, $T_H=\frac{1}{8\pi M}$,
for the ordinary Schwarzschild black hole.

\section{Entropy of noncommutative inspired Schwarzschild black hole}
In this section, we discuss the entropy of the
NC inspired Schwarzschild black hole and its difference after and before a massive particle's emission.
In the case of massless tunneling~\cite{Wilczek99}, ${\rm Im} S = -\frac{1}{2} \Delta S_{\rm BH}$. But in our case, this equality no longer holds.
We calculate the entropy of the NC inspired Schwarzschild black hole by using the first law of thermodynamics, $dM=TdS$,
\begin{eqnarray}
S_{\rm BH}&=&\int^M_{M_0}\frac{dM'}{T_H}\nonumber\\
&=&4\pi\left[M^2-\frac{3}{2}\th\right]{\rm erf}\left(\frac{M}{\sqrt\th}\right)+12\sqrt{\pi\th}\, M e^{-\frac{M^2}{\th}}\nonumber\\
& &-4\pi\left[M^2_0-\frac{3}{2}\th\right]{\rm erf}\left(\frac{M_0}{\sqrt\th}\right)-12\sqrt{\pi\th}\, M_0 e^{-\frac{M^2_0}{\th}},
\label{entro}
\end{eqnarray}
where the temperature given by eq.~(\ref{temperature}) has been used. Note that the lower integration limit
cannot be set to zero but to the mass of the extreme black hole $M_0$ because the NC inspired Schwarzschild black hole can exist only for $M \ge M_0$.
This is a basic feature which we should consider in the calculation of the entropy, however,
it was unnoticed by some literature~\cite{NCbhmassless,NCbhmassive}.
The entropy can reduce to the standard Bekenstein-Hawking entropy, $S_{\rm BH} = 4\pi M^2$,
for the ordinary Schwarzschild black hole in the commutative limit $\th \rightarrow 0$, in which case no extreme black holes exist.
With eq.~(\ref{entro}) we obtain the difference of the entropy after and before the emission,
\begin{eqnarray}
\Delta S_{\rm BH}&=& S_{\rm BH}(M-\omega)-S_{\rm BH}(M)\nonumber\\
&=&4\pi\left[(M-\omega)^2-\frac{3}{2}\th\right]{\rm erf}\left(\frac{M-\omega}{\sqrt\th}\right)
+12\sqrt{\pi\th}\,(M-\omega )e^{-\frac{(M-\omega)^2}{\th}}\nonumber\\
& & -4\pi\left[M^2-\frac{3}{2}\th\right]{\rm erf}\left(\frac{M}{\sqrt\th}\right)-12\sqrt{\pi\th}\,M e^{-\frac{M^2}{\th}},
\end{eqnarray}
which is independent of the mass of the extreme black hole.
Comparing it with eq.~(\ref{imaction}), we see ${\rm Im} S\neq -\frac{1}{2} \Delta S_{\rm BH}$. When we take the limit $m \rightarrow 0$,
which corresponds to the case of massless particle's tunneling, the tunneling probability equals the exponent of the difference of entropy,
$e^{-2 {\rm Im} S}= e^{\Delta S_{\rm BH}}$. As a consequence, our result is consistent with that of ref.~\cite{Wilczek99}.

\section{Conclusion}
In this paper, we analyze the tunneling process of massive particles from the NC inspired Schwarzschild black hole following the way given
in our previous work~\cite{Mxz2010} which is based on the Parikh-Wilczek method~\cite{Wilczek99}.
We provide some novel and previously unnoticed properties for
the noncommutative inspired Schwarzschild black hole, such as:
(i) We generalize the Parikh-Wilczek method to massive tunneling particles in methodology;
(ii) The tunneling rate of massive particles is suppressed due to the mass, and the effective radiation temperature
is modified by the noncommutativity but recovers the standard Hawking temperature in the commutative limit,
and eqs.~(\ref{imaction}) and (\ref{entropy}) imply that the radiation is not precisely thermal, as pointed out in ref.~\cite{Wilczek99};
(iii) The well-known conclusion that the tunneling rate equals the exponent of the difference of the entropy does not hold in
the massive case but it can be recovered in the zero mass limit, which coincides with that of ref.~\cite{Wilczek99}.
This inequality may imply that there exist correlations between tunneling particles.
To see this,
as done in refs.~\cite{NCbhmassless,NCbhmassive}, we can compare the probability of tunneling of two particles of energies (masses)
$\om_1$ ($m_1$) and $\om_2$ ($m_2$) with the probability of tunneling of one particle with the energy (mass) summation $\om_1+\om_2$ ($m_1+m_2$).
We obtain an inequality,
\begin{equation}
\ln\Gamma_{\omega_1, m_1} + \ln\Gamma_{\omega_2, m_2} \neq \ln \Gamma_{\omega_1+\omega_2, m_1+m_2},
\end{equation}
for the massive case, but an equality, $\ln\Gamma_{\omega_1} + \ln\Gamma_{\omega_2} = \ln \Gamma_{\omega_1+\omega_2}$,
for the massless case $m_1=m_2=0$.
Therefore, the radiation may also contain information beyond the remnant for the massive case, which deserves further investigations.

\vskip 10mm

\section*{Acknowledgments}
Y-GM would like to thank J.-X. Lu of the Interdisciplinary Center for Theoretical Study (ICTS),
University of Science and Technology of China (USTC) for warm hospitality.
This work is supported by the National Natural
Science Foundation of China under the projects Nos.11175090 and 10675061, and by the Fundamental Research Funds for the Central Universities
No.65030021.
At last, the authors would like to thank the anonymous referees for their helpful comments which indeed improve this paper greatly.

\end{document}